\documentclass[12pt,onecolumn,draftcls]{IEEEtran}
\usepackage[T1]{fontenc}
\usepackage[latin9]{inputenc}
\usepackage{color}
\usepackage{array}
\usepackage{float}
\usepackage{amsmath}
\usepackage{amssymb}
\usepackage{graphicx}
\usepackage{setspace}
\usepackage{wasysym}
\usepackage[unicode=true,
 bookmarks=true,bookmarksnumbered=true,bookmarksopen=true,bookmarksopenlevel=1,
 breaklinks=false,pdfborder={0 0 0},pdfborderstyle={},backref=false,colorlinks=false]
 {hyperref}
\hypersetup{pdftitle={Your Title},
 pdfauthor={Your Name},
 pdfpagelayout=OneColumn, pdfnewwindow=true, pdfstartview=XYZ, plainpages=false}

\makeatletter

\providecommand{\tabularnewline}{\\}
\newcommand{\lyxdot}{.}

\floatstyle{ruled}
\newfloat{algorithm}{tbp}{loa}
\providecommand{\algorithmname}{Algorithm}
\floatname{algorithm}{\protect\algorithmname}

\usepackage[caption=false,font=footnotesize]{subfig}
\usepackage{algorithmic}

\makeatother

\begin{document}
\title{Multiband Delay Estimation for Localization Using a Two-Stage Global
Estimation Scheme}
\author{\singlespacing{}{\normalsize{}Yubo Wan, }\textit{\normalsize{}Graduate Student Member,
IEEE}{\normalsize{}, An Liu, }\textit{\normalsize{}Senior Member,
IEEE}{\normalsize{}, Qiyu Hu, }\textit{\normalsize{}Graduate Student
Member, IEEE}{\normalsize{}, Mianyi Zhang, and Yunlong Cai,}\textit{\normalsize{}
Senior Member, IEEE}{\normalsize{}}\thanks{This work was supported in part by National Science Foundation of
China (No.62071416), and in part by Huawei Techologies Co., Ltd. (Corresponding
author: An Liu.)\protect \\
Yubo Wan, An Liu, Qiyu Hu, Mianyi Zhang, and Yunlong Cai are with
the College of Information Science and Electronic Engineering, Zhejiang
University, Hangzhou 310027, China (email: wanyb@zju.edu.cn; anliu@zju.edu.cn;
qiyhu@zju.edu.cn; mianyi\_zhang@zju.edu.cn; ylcai@zju.edu.cn).}}

\maketitle
\vspace{-0.5in}

\begin{abstract}
The time of arrival (TOA)-based localization techniques, which need
to estimate the delay of the line-of-sight (LoS) path, have been widely
employed in location-aware networks. To achieve a high-accuracy delay
estimation, a number of multiband-based algorithms have been proposed
recently, which exploit the channel state information (CSI) measurements
over multiple non-contiguous frequency bands. However, to the best
of our knowledge, there still lacks an efficient scheme that fully
exploits the multiband gains when the phase distortion factors caused
by hardware imperfections are considered, due to that the associated
multi-parameter estimation problem contains many local optimums and
the existing algorithms can easily get stuck in a ``bad'' local
optimum. To address these issues, we propose a novel two-stage global
estimation (TSGE) scheme for multiband delay estimation. In the coarse
stage, we exploit the group sparsity structure of the multiband channel
and propose a Turbo Bayesian inference (Turbo-BI) algorithm to achieve
a good initial delay estimation based on a coarse signal model, which
is transformed from the original multiband signal model by absorbing
the carrier frequency terms. The estimation problem derived from the
coarse signal model contains less local optimums and thus a more stable
estimation can be achieved than directly using the original signal
model. Then in the refined stage, with the help of coarse estimation
results to narrow down the search range, we perform a global delay
estimation using a particle swarm optimization-least square (PSO-LS)
algorithm based on a refined multiband signal model to exploit the
multiband gains to further improve the estimation accuracy. Simulation
results show that the proposed TSGE significantly outperforms the
benchmarks with comparative computational complexity.
\end{abstract}


\section{Introduction}

Recently, the wireless-based localization has attracted great attention
due to its adaptability to the existing wireless infrastructure and
capacity for assisting communications \cite{survey_localization1,survey_localization2}.
It has been widely employed in multisensory extended reality (XR)
\cite{XR}, smart transportation \cite{smarttrans1,smarttrans2},
and connected robotics and autonomous systems (CRAS) \cite{CRAS}.
These applications generally require submeter-level localization accuracy
and even centimeter-level localization accuracy. Initially, the global
navigation satellite system (GNSS) has been employed to provide location
services, but it has low accuracy in indoor environments \cite{GPS}.
Thus, the cellular-based localization and wireless local area networks
(WLAN)-based localization have been developed as alternatives to GNSS
\cite{wireless_loc1,wireless_locTOA2,wireless_locTOA3,Wifi_local1}.
These approaches estimate the positions of agents by utilizing the
features inferred from the radio frequency (RF) signals, which include
time of arrival (TOA), angle of arrival (AoA), angle of departure
(AoD), time difference of arrival (TDOA), and received signal strength
(RSS). In particular, the TOA-based localization is a widely studied
wireless localization method \cite{wireless_locTOA2,wireless_locTOA3,Wifi_local1,wireless_locTOA4,wireless_locTOA5},
which estimates the distance by multiplying the delay of the line-of-sight
(LoS) path with the light speed.

\begin{figure}[t]
\centering{}\includegraphics[width=10cm]{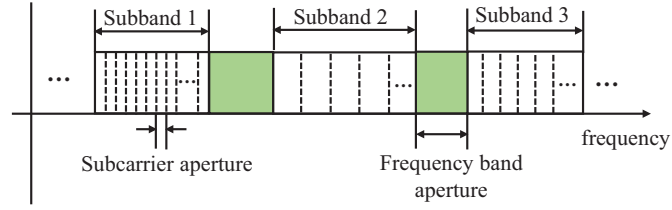}\caption{\label{fig:The-distribution-of}An illustration of subcarrier and
frequency band apertures.}
\end{figure}

However, the estimation accuracy of the multipath channel delays is
limited by the bandwidth of the transmitted signal. To address this
issue, the multiband delay estimation schemes have been proposed in
\cite{XuHuilin1,nsdi,CS2019,CS2020,ESPRIT1,ESPRIT2,MUSIC_1,LTESAGE},
which make use of the channel state information (CSI) measurements
across multiple frequency bands to obtain high accuracy delay estimation.
Compared to single band delay estimation, multiband delay estimation
can obtain extra multiband gains, which consist of two parts: (i)
Multiband CSI samples lead to more subcarrier apertures gain; (ii)
Frequency band apertures gain brought by the difference of carrier
frequency between subbands \cite{ESPRIT2}. The subcarrier apertures
and the frequency band apertures are shown in Fig. \ref{fig:The-distribution-of}.
As can be seen, the spectrum resource used for localization is non-contiguous,
that consists of a number of subbands. The green regions are the frequency
subbands allocated to other applications (e.g., wireless communication)
and thus are unable to utilize for localization. Despite the existence
of the apertures gains, the multiband delay estimation method still
faces new challenges. One challenge is the phase distortion in the
channel frequency response (CFR) samples caused by hardware imperfections
\cite{nonidealfoctor1,nonidealfoctor2,nonidealfoctor3}, which has
a severe effect on delay estimation and needs to be calibrated. The
other challenge is that the high frequency carrier in the CFR samples
leads to a violent oscillation phenomenon of the likelihood function,
which causes numerous bad local optimums \cite{ESPRIT1,ESPRIT2}.
Therefore, it is very difficult to exploit the frequency band apertures
gain. Some related works are summarized below.

\textbf{Maximum likelihood (ML) based methods:} The traditional approaches
to estimate the delay parameters are ML based estimation methods,
of which the space-alternating generalized expectation-maximization
(SAGE) algorithm stands out for its versatility and robustness in
harsh multipath environments \cite{SAGE_base}. In \cite{LTESAGE},
a multiband TOA estimation method has been carried out by the SAGE
algorithm for long term evolution (LTE) downlink systems, which provides
a reduced standard deviation for the delay estimation. However, the
authors in \cite{ML_localoptimun} pointed out that the ML based methods
tend to converge to a local optimum and need to find the global optimum
in a large searching space. To reduce the computational complexity,
the authors in \cite{XuHuilin1} proposed a low-complexity approach,
which recovers the channel impulse response (CIR) with equally spaced
taps and approximately estimates the first path by mitigating the
energy leakage. However, this method estimates the delay of the first
path coarsely and results in a limited performance improvement.

\textbf{Subspace based estimation methods:} There have been works
attempting to solve the multiband delay estimation problems by using
subspace based methods, such as multiple signal classification (MUSIC)
\cite{MUSIC_1} and estimating signal parameters via rotational invariance
techniques (ESPRIT) \cite{ESPRIT1,ESPRIT2}. In \cite{MUSIC_1}, the
classical MUSIC algorithm has been applied to delay estimation. However,
this approach simply exploits the subcarrier apertures gain while
the frequency band apertures gain is not exploited, which results
in a performance loss. The authors in \cite{ESPRIT1,ESPRIT2} employed
the multiple shift-invariance structure in the multiband channel measurements
and thus the proposed algorithm achieves a high accuracy delay estimation.
Nevertheless, the subcarrier spacing of different subbands is assumed
to be equivalent, which restricts its applications for practical systems.
Moreover, the subspace based methods generally require multiple snapshots
of orthogonal frequency division multiplexing (OFDM) pilot symbols
to guarantee the performance \cite{MUSICbase}, which consumes lots
of pilot resources.

\textbf{Compressed sensing (CS) based methods:} In an indoor environment,
the CIR is sparse since it consists of only a small number of paths.
Motivated by the sparsity of CIR over the delay domain, many state-of-the-art
delay estimation algorithms have been proposed based on CS methods
\cite{nsdi,CS2019,CS2020}. In \cite{nsdi,CS2019}, the authors formulated
$l_{1}$-norm minimization problems for capturing the signal sparsity.
In \cite{CS2020}, orthogonal matching pursuit (OMP) methods have
been used for recovering the sparse CIR. However, these approaches
have the problem of energy leakage resulting from basis mismatch \cite{FDD2D},
which require dense grids and have high computational complexity.

In the aforementioned studies, on one hand, the accuracy of most algorithms
is limited, since the associated multi-parameter estimation problem
contains many bad local optimums caused by high frequency carrier
terms and frequency band apertures. Though some other works have eliminated
the oscillation, they have not exploited all apertures gains in the
multiband CSI measurements \cite{nsdi,MUSIC_1}, e.g., most works
only exploit the subcarrier apertures gain due to the difficulty of
exploiting the frequency band apertures gain. On the other hand, many
works have not considered the imperfect phase distortion factors caused
by receiver timing offset or phase noise in practical systems \cite{XuHuilin1,ESPRIT1,MUSIC_1,XuHuilin2}.
Though some studies in \cite{nsdi,CS2020,ESPRIT2} have considered
the phase distortion, the calibration of the phase distortions with
extra information is required via a handshaking procedure under the
assumption of channel reciprocity. However, this assumption is restrictive
and the transceiver needs to have the ability of Tx/Rx switching.
Furthermore, in the handshaking procedure, the phase lock loop (PLL)
must keep inlock to ensure that the phase offset remain an unchanged
absolute value, which is difficult to achieve in practice.

In this paper, we consider a TOA-based localization system using OFDM
training signals over multiple frequency subbands. Then, a novel two-stage
global estimation (TSGE) scheme is proposed to fully exploit all the
multiband gains, where we consider all the phase distortion factors
and calibrate them implicitly without using extra handshaking procedures.

Specifically, in Stage 1, we build a coarse signal model, in which
the high frequency carriers are all absorbed for eliminating the oscillation
of the likelihood function. Although the coarse estimation algorithm
derived from the coarse signal model can only exploit the subcarrier
apertures gain, it does not get stuck in bad local optimums and thus
can provide a much more stable delay estimation to narrow down the
search range for the global delay estimation in the refined stage.
Then, we provide a sparse representation for multiband CFR, where
we adopt a common support based sparse vector to capture the group
sparsity structure in the multiband channel over the delay domain.
Based on this model, a Turbo Bayesian inference (Turbo-BI) algorithm
is proposed for channel parameter estimation (including the delay
parameter). Compared to the CS-based delay estimation methods in \cite{nsdi,CS2019,CS2020},
our proposed algorithm achieves higher estimation accuracy with lower
computational complexity. It is because we adopt a dynamic grid adjustment
strategy and we do not need a very dense grid.

In Stage 2, with the help of prior information passed from Stage 1,
we perform a finer estimation based on a refined signal model. A higher
estimation accuracy than that in Stage 1 can be guaranteed since this
signal model contains the structure of frequency band apertures. However,
the refined signal model leads to a multi-dimensional non-convex likelihood
function that has many bad local optimums, which makes it extremely
difficult to find the global optimum and fully exploit the frequency
band apertures gain. For utilizing this apertures gain and the prior
information properly, we adopt a global search algorithm based on
the particle swarm optimization (PSO) to find a good solution for
the non-convex optimization of the multi-dimensional likelihood function.
In particular, the coarse estimation results from Stage 1 can be utilized
for determining the particle search range, which can reduce the search
complexity significantly. For further reducing the search space and
improving the estimation accuracy, we employ primal-decomposition
theory to decouple the objective function and get a least square (LS)
solution for channel coefficients. Then, the dimension of the search
space can be reduced by eliminating the channel coefficients in the
primal optimization problem. The main contributions are summarized
below.
\begin{itemize}
\item A novel TSGE scheme is proposed for obtaining multiband gains with
acceptable complexity, which includes a coarse estimation stage and
a refined estimation stage based on a two-stage signal model.
\item In Stage 1, we set up a common support based probability model, which
employs the group sparsity structure of the multiband channel. Then,
a Turbo-BI algorithm is proposed for delay estimation.
\item In Stage 2, we propose a PSO-LS algorithm based on PSO and primal-decomposition
theory, which reduces the dimension of the search space and significantly
improves the estimation accuracy.
\end{itemize}
Finally, extensive simulation results are presented to validate the
effectiveness of the proposed scheme and we show that it can achieve
superior delay estimation accuracy as compared to the baseline schemes.

The rest of this paper is organized as follows. In Section \ref{sec:System Model},
we describe the system model and introduce the phase distortion factors.
In Section \ref{sec:Two-Stage-Multiband-Delay}, we formulate the
two-stage signal model and outline the TSGE scheme. Sections \ref{sec:Turbo-BI-Algorithm-in}
and \ref{sec:PSO-LS-algorithm-in} present the Turbo-BI algorithm
and the PSO-LS algorithm in Stage 1 and Stage 2, respectively. In
Section \ref{sec:Simulation-Results}, numerical results are presented
and finally Section \ref{sec:Conclusion} concludes the paper.

\textit{Notations}: Bold upper (lower)-case letters are used to define
matrices (column vectors). In particular, bold letters indexed by
subscript $m$ denote vectors or matrices corresponding to the $m$-th
subband. $\mathbf{I}$ denotes an identity matrix, $\delta\left(\cdot\right)$
denotes the Dirac's delta function, $\mathrm{diag}\left(\cdot\right)$
constructs a diagonal matrix from its vector argument, and $\|\cdot\|$
denotes the Euclidean norm of a complex vector. For a matrix $\mathbf{A}$,
$\mathbf{A}^{T},\mathbf{A}^{H},\mathbf{A}^{-1},\mathrm{tr}(\mathbf{A})$
represent a transpose, complex conjugate transpose, inverse, and trace
of a matrix, respectively. For a scalar $a$, $a^{*}$ denotes the
conjugate of a scalar. The notation $\mathbb{R}^{\textrm{+}}$ represents
the strictly positive real number and $\mathcal{CN}(\mathbf{x};\mathrm{\boldsymbol{\mu}},\boldsymbol{\Sigma})$
denotes a complex Gaussian normal distribution corresponding to variable
$\mathbf{x}$ with mean $\boldsymbol{\mu}$ and covariance matrix
$\boldsymbol{\Sigma}$.

\section{System Model\label{sec:System Model}}

\begin{figure}[t]
\centering{}\includegraphics[width=10cm]{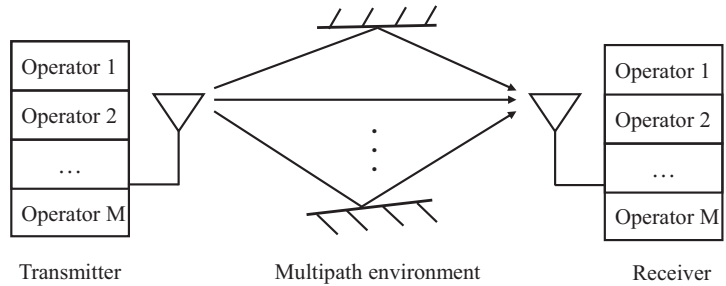}\caption{\label{fig:The-multiband-OFDM}An illustration of the multiband OFDM
system.}
\end{figure}

As shown in Fig. \ref{fig:The-multiband-OFDM}, we consider a single-input
single-output (SISO) multiband system which employs OFDM training
signals over $M$ frequency subbands. The multiband system consists
of $M$ non-overlapping single band OFDM subsystems, where the $m$-th
subband is allocated to the $m$-th operator. Assume that each frequency
subband has $N_{m}$ orthogonal subcarriers with subcarrier spacing
$f_{s,m}$ and the carrier frequency of subband $m$ is denoted as
$f_{c,m}$. Then, the continuous-time CIR $h\left(t\right)$ can be
written as
\begin{equation}
h(t)=\sum_{k=1}^{K}\alpha_{k}\delta\left(t-\tau_{k}\right),
\end{equation}
where $K$ denotes the number of multipath components between the
transmitter and the receiver, $\alpha_{k}\in\mathbb{C}$ and $\tau_{k}\in\mathbb{R^{\textrm{+}}}$
denote the complex path gain and the delay of the $k$-th path, respectively.
The delays are sorted in an increasing order, i.e., $\tau_{k-1}<\tau_{k}$,
$k=2,...,K$, and $\tau_{1}$ is the LoS path which needs to be estimated
for localization. We assume that the complex path gain and delay parameters
are independent of the frequency subbands. Then, via a Fourier transform
of the CIR as in \cite{CS2019,ESPRIT2}, the CFR samples can be expressed
as
\begin{equation}
\tilde{h}_{m,n}=\sum_{k=1}^{K}\alpha_{k}e^{-j2\pi f_{m,n}\tau_{k}},\label{eq:CFR}
\end{equation}
where $f_{m,n}=f_{c,m}+n_{m}f_{s,m}$, $m=1,...,M$, $n_{m}\in\mathcal{N_{\mathit{m}}}\triangleq\left\{ -\frac{N_{m}}{2},...,\frac{N_{m}}{2}-1\right\} $.
With a slight abuse of notation, we use $n$ instead of $n_{m}$ in
the following equations. We assume that $N_{\mathit{m}},\forall m$
is an even number without loss of generality, and denote $N=N_{1}+\ldots+N_{M}$
as the number of CFR samples over all subbands.

Apparently, the CFR exhibits sparsity over delay domain when $K$
is small, which will be exploited in our proposed Turbo-BI algorithm.
Then, during the period of a single OFDM symbol, the discrete-time
received signal model can be written as \cite{CS2019,ESPRIT2}
\begin{equation}
y_{m,n}=\sum_{k=1}^{K}\alpha_{k}e^{-j2\pi\left(f_{c,m}+nf_{s,m}\right)\tau_{k}}e^{-j2\pi nf_{s,m}\delta_{m}}e^{j\varphi_{m}}s_{m,n}+w_{m,n},\label{eq:original_signal}
\end{equation}
where $w_{m,n}$ is the $n$-th element of the additive white Gaussian
noise (AWGN) vector $\boldsymbol{w}_{m}\in\mathbb{C}^{N_{m}\times1}$,
following the distribution $\mathcal{C}\mathcal{N}\left(0,\sigma_{ns}^{2}\mathbf{I}\right)$.
$s_{m,n}$ denotes a known training symbol over the $n$-th subcarrier
of subband $m$ and we assume $s_{m,n}=1,\forall m,n$ for simplicity.
The parameter $\varphi_{m}$ and $\delta_{m}$ represent the phase
distortion factors caused by random phase offset and receiver timing
offset \cite{nonidealfoctor1,nonidealfoctor2,nonidealfoctor3}, respectively.
In practice, the receiver timing offset $\delta_{m}$ is often within
a small range and thus we assume that $\delta_{m},\forall m$ follows
a prior distribution $p\left(\delta_{m}\right)\sim\mathcal{N}\left(0,\sigma_{p}^{2}\right)$
\cite{nonidealfoctor3}, where $\sigma_{p}$ is the timing synchronizaton
error.

However, the global delay estimation problem will be intractable if
we directly use the signal model (\ref{eq:original_signal}) due to
the huge search space of the multi-dimensinal parameters and the existence
of many local optimums in the likelihood function, which thus motivates
the set up of the novel two-stage signal model and the associated
two-stage global estimation scheme.

\section{\label{sec:Two-Stage-Multiband-Delay}Two-Stage Global Estimation
Scheme for Multiband Delay Estimation}

In this section, we first explain why we cannot use the original signal
model for delay estimation based on the likelihood function analysis,
which motivates us to introduce the proposed two-stage signal model.
Then, based on the two-stage signal model, we provide the outline
of the proposed TSGE scheme.

\subsection{\label{subsec:Two-Stage-Signal-Model}Two-Stage Signal Model}

It is difficult to directly use the received signal model (\ref{eq:original_signal})
for delay estimation. On one hand, the original model (\ref{eq:original_signal})
exists inherent ambiguity. Specifically, for an arbitrary constant
$c$, if we substitute two sets of variables $(\left|\alpha_{k}\right|e^{j\angle\alpha_{k}},\phi_{m})$
and $(\left|\alpha_{k}\right|e^{j(\angle\alpha_{k}+c)},\phi_{m}-c)$
into equation (\ref{eq:original_signal}), the same observation result
will be obtained. It indicates that the parameters $(\alpha_{k},\phi_{m})$
are ambiguous, which leads to the difficulty of delay estimation.

\begin{figure}[t]
\centering{}\includegraphics[width=10cm]{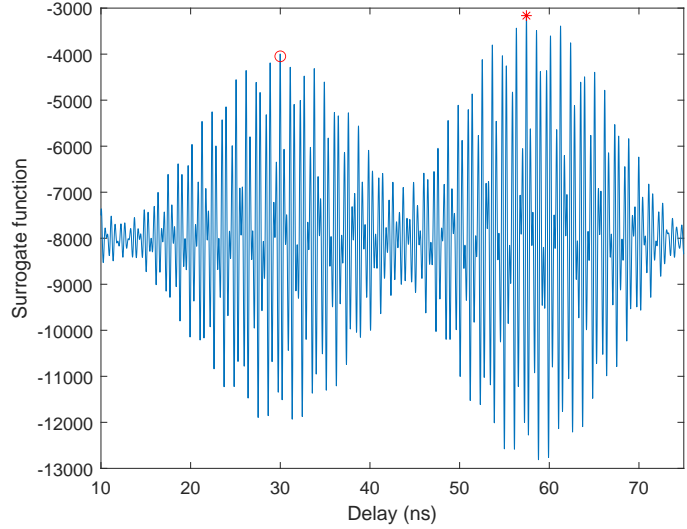}\caption{\label{fig:Likehood-original}An illustration of the likelihood function
based on signal model (\ref{eq:original_signal}).}
\end{figure}

On the other hand, due to the high frequency carrier $f_{c,m}$, the
original signal model (\ref{eq:original_signal}) leads to a multimodal
non-convex likelihood function that has many sidelobes. Consequently,
it is extremely difficult to find the global optimum. In the worst
case, the point of true values may fall into sidelobes, which will
inevitably result in a large estimation error. To clarify this problem,
we plot a likelihood function curve for the one-dimensional problem
of estimating the LoS path delay only in Fig. \ref{fig:Likehood-original},
where the red circle marks the point of the true values of LoS path
delay and the red star marks the point of global optimum. As can be
seen, the likelihood function fluctuates frequently, which makes it
intractable to find the global optimum. Moreover, the point of true
values is not in the mainlobe, at which the global optimum locates.
Even though we try our best to find the global optimum, an absolute
estimation error about $27$ ns is still inevitable.

Therefore, we build new signal models without inherent ambiguity and
the probability that the point of true values is in the region of
the mainlobe is much higher. Moreover, the signal models should reserve
the structure of frequency band apertures and subcarrier apertures.
Motivated by the above facts, we propose a two-stage signal model,
which is transformed from the original signal model (\ref{eq:original_signal})
by absorbing different frequency/phase terms into the complex gain.

\begin{figure}[t]
\centering{}\includegraphics[width=10cm]{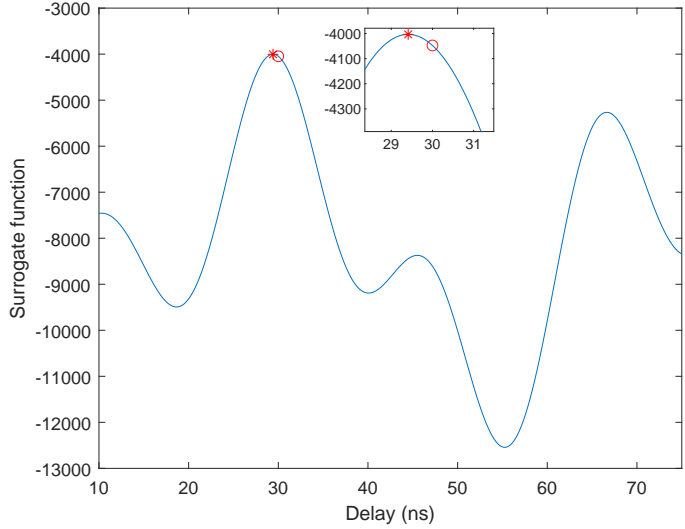}\caption{\label{fig:Likehood-coarse}An illustration of the likelihood function
based on signal model (\ref{eq:coarse_signal}).}
\end{figure}

\subsubsection{Coarse Signal Model}

\begin{equation}
y_{m,n}=\sum_{k=1}^{K}\alpha_{k,m}e^{-j2\pi nf_{s,m}\tau_{k}}e^{-j2\pi nf_{s,m}\delta_{m}}+w_{m,n},\label{eq:coarse_signal}
\end{equation}
where $\alpha_{k,m}=\alpha_{k}e^{j\varphi_{m}}e^{-j2\pi f_{c,m}\tau_{k}},\forall k,m$.
In signal model (\ref{eq:coarse_signal}), we absorb the terms of
random phase offset $e^{j\varphi_{m}}$ and carrier phase $e^{-j2\pi f_{c,m}\tau_{k}}$
into $\alpha_{k}$. Signal model (\ref{eq:coarse_signal}) becomes
unambiguous according to this equivalent transformation and all subbands
share the common sparse delay domain. As shown in Fig. \ref{fig:Likehood-coarse},
we depict the likelihood function based on signal model (\ref{eq:coarse_signal})
with the same parameter values as Fig. \ref{fig:Likehood-original}.
The likelihood function no longer frequently fluctuates and the point
of true values locates at the mainlobe region. In this case, we can
exploit the subcarrier apertures gain and estimate delay parameters
without ambiguity. This helps to achieve a much more stable delay
estimation in the coarse estimation stage whose primary purpose is
to narrow down the search range to a relatively small region with
high stability (probability). However, the estimation accuracy is
limited, since we have absorbed the carrier phase term and thus cannot
exploit the frequency band apertures gain, which motivates the next
refined estimation signal model (\ref{eq:refined_signal}) in Stage
2.

\begin{figure}[t]
\centering{}\includegraphics[width=10cm]{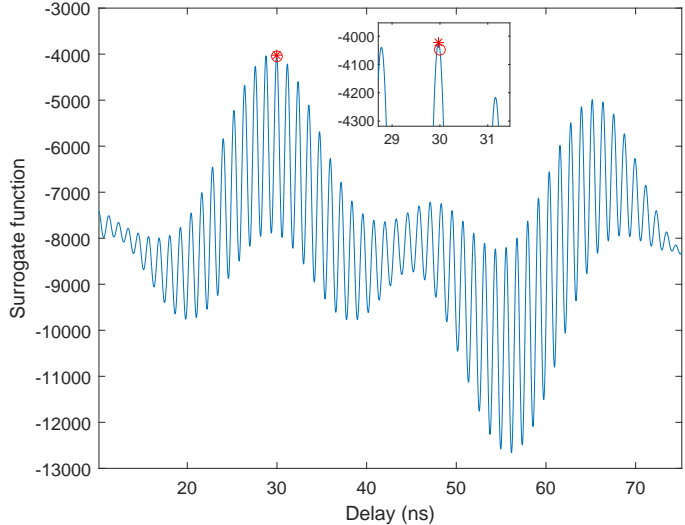}\caption{\label{fig:Likehood-jinggu}An illustration of the likelihood function
based on signal model (\ref{eq:refined_signal}).}
\end{figure}

\subsubsection{Refined Signal Model}

\begin{equation}
y_{m,n}=\sum_{k=1}^{K}\alpha_{k}^{\prime}e^{-j2\pi f_{c,m}^{\prime}\tau_{k}}e^{-j2\pi nf_{s,m}\tau_{k}}e^{-j2\pi nf_{s,m}\delta_{m}}e^{j\varphi_{m}^{\prime}}+w_{m,n},\label{eq:refined_signal}
\end{equation}
where $f_{c,m}^{\prime}=f_{c,m}-f_{c,1}$, $\alpha_{k}^{\prime}=\alpha_{k}e^{j\varphi_{1}}e^{-j2\pi f_{c,1}\tau_{k}}$,
$\varphi_{m}^{\prime}=\varphi_{m}-\varphi_{1}$,$\forall k,m.$ In
the signal model (\ref{eq:refined_signal}), we absorb the random
phase offset, $e^{j\varphi_{1}}$, and carrier phase of the first
frequency subband, $e^{-j2\pi f_{c,1}\tau_{k}}$, into $\alpha_{k}$
and reserve the residual term, e.g., $e^{-j2\pi\left(f_{c,2}-f_{c,1}\right)\tau_{k}}$
and $e^{j\left(\varphi_{2}-\varphi_{1}\right)}$ when $m=2$, in (\ref{eq:refined_signal}).
Compared to (\ref{eq:coarse_signal}), signal model (\ref{eq:refined_signal})
has the extra structure of frequency band apertures, $e^{-j2\pi f_{c,m}^{\prime}\tau_{k}}$,
and residual random phase offset, $e^{j\varphi_{m}^{\prime}}$. Fig.
\ref{fig:Likehood-jinggu} illustrates the likelihood function based
on signal model (\ref{eq:refined_signal}), where the likelihood function
fluctuates less than that in Fig. \ref{fig:Likehood-original} and
the point of true values is now in the mainlobe region. Moreover,
we observe that the mainlobe is sharper than that in Fig. \ref{fig:Likehood-coarse}
due to the the existing of frequency band apertures, which leads to
a potential performance improvement. However, the frequency band apertures
also cause numerous bad local optimums in the likelihood function.
How to find the global optimum with low-complexity is a challenging
problem. To solve this problem, we propose the PSO-LS algorithm later.

In summary, both two signal models are essential. In Stage 1 based
on signal model (\ref{eq:coarse_signal}), we exploit the subcarrier
apertures and give a stable coarse estimation as the initial result
to Stage 2. Then, in Stage 2, signal model (\ref{eq:refined_signal})
is used for providing a refined estimation by exploiting both the
subcarrier apertures gain and the frequency band apertures gain.

\subsection{Outline of the TSGE Scheme}

Based on the two-stage signal model, the TSGE scheme is depicted as
follows:
\begin{itemize}
\item Stage 1: We set up the coarse signal model (\ref{eq:coarse_signal})
and perform an initial delay estimation using the proposed Turbo-BI
algorithm. By doing this, we exploit the channel sparsity over delay
domain and the subcarrier apertures gain. Then, we provide the estimation
result to Stage 2.
\item Stage 2: Based on the coarse estimation result from Stage 1 and the
refined signal model (\ref{eq:refined_signal}), a more refined delay
estimation is performed. To fully exploit subcarrier and frequency
band apertures gains and overcome the difficulty of finding the global
optimum as shown in Fig. \ref{fig:Likehood-jinggu}, we propose the
PSO-LS algorithm.
\end{itemize}
The overall algorithm is summarized as in Algorithm \ref{alg:TSMBDE-scheme}.
The details of the Turbo-BI and PSO-LS algorithms are presented in
Section \ref{sec:Turbo-BI-Algorithm-in} and \ref{sec:PSO-LS-algorithm-in},
respectively.

\begin{algorithm}[tbh]
{\small{}\caption{\label{alg:TSMBDE-scheme}TSGE scheme}
}{\small\par}

\textbf{Input:} CFR samples $y_{m,n},\forall m,n$.

\textbf{Output:} The delay estimation result from Stage 2.

\begin{algorithmic}[1]

\STATE \textbf{Stage 1:}

\STATE \textbf{\%Coarse estimation}

\STATE Construct the coarse signal model (\ref{eq:coarse_signal}).

\STATE Common support based sparse representation for the coarse
signal model (\ref{eq:coarse_signal}).

\STATE Perform Turbo-BI algorithm.

\STATE Pass the coarse estimation result to Stage 2.

\STATE \textbf{Stage 2:}

\STATE \textbf{\%Refined estimation}

\STATE Construct the refined signal model (\ref{eq:refined_signal}).

\STATE Perform primal-decomposition for problem (\ref{eq:P2}).

\STATE Perform PSO-LS algorithm with the the initial particles generated
using the coarse estimation in Stage 1.

\end{algorithmic}
\end{algorithm}

\section{\label{sec:Turbo-BI-Algorithm-in}Turbo-BI Algorithm in Stage 1}

\subsection{Common Support Based Sparse Representation}

We first describe the sparse representation over delay domain for
the coarse signal model (\ref{eq:coarse_signal}), which is a necessary
step before employing the sparse recovery methods, e.g., Turbo-BI
algorithm. One commonly used method is to define a uniform grid $\mathcal{D}=\left\{ \overline{d}_{1},\ldots,\overline{d}_{L}\right\} $
of $L$ ($L\gg K$) delay points over $\left[0,T_{d}\right]$ ($T_{d}$
denotes an upper bound for the maximum delay spread). If all the true
delay values exactly lie in the discrete set $\mathcal{D}$, we can
reformulate the signal model (\ref{eq:coarse_signal}) as:
\begin{equation}
\left[\begin{array}{c}
\mathbf{y}_{1}\\
\vdots\\
\mathbf{y}_{M}
\end{array}\right]=\left[\begin{array}{lll}
\mathbf{S}_{1}\\
 & \ddots\\
 &  & \mathbf{S}_{M}
\end{array}\right]\cdot\left[\begin{array}{ccc}
\mathbf{A}_{1}\\
 & \ddots\\
 &  & \mathbf{A}_{M}
\end{array}\right]\cdot\left[\begin{array}{c}
\mathbf{x}_{1}\\
\vdots\\
\mathbf{x}_{M}
\end{array}\right]+\left[\begin{array}{c}
\boldsymbol{w}_{1}\\
\vdots\\
\boldsymbol{w}_{M}
\end{array}\right],\label{eq:ULAsignal}
\end{equation}
where $\mathbf{y}_{m}=[y_{m,-\frac{N_{m}}{2}},\cdots,y_{m,\frac{N_{m}}{2}-1}]^{T}\in\mathbb{C}^{N_{m}\times1}$,
$\mathbf{A}_{m}=[\mathbf{a}_{m}(\overline{d}_{1}),\mathbf{a}_{m}(\overline{d}_{2}),\cdots,\mathbf{a}_{m}(\overline{d}_{L})]\in\mathbb{C}^{N_{m}\times L}$
denotes the basis matrix, $\mathbf{a}_{m}(\overline{d}_{l})=[e^{-j2\pi(-\frac{N_{m}}{2})f_{s,m}\overline{d}_{l}},\ldots,e^{-j2\pi(\frac{N_{m}}{2}-1)f_{s,m}\overline{d}_{l}}]^{T}\in\mathbb{C}^{N_{m}\times1}$
is a linear steering vector, $\mathbf{S}_{m}=\mathrm{diag}(e^{-j2\pi(-\frac{N_{m}}{2})f_{s,m}\delta_{m}},\ldots,$$e^{-j2\pi(\frac{N_{m}}{2}-1)f_{s,m}\delta_{m}})\in\mathbb{C}^{N_{m}\times N_{m}}$,
and $\mathbf{x}_{m}\in\mathbb{C}^{L\times1}$ is a sparse vector whose
non-zero elements correspond to the true delays. For example, if the
$l$-th element of $\mathbf{x}_{m}$ denoted by $x_{m,l}$ is non-zero
and the corresponding true delay is $\tau_{\hat{k}}$, then we have
$\overline{d}_{l}=\tau_{\hat{k}}$ and $x_{m,l}=\alpha_{\hat{k},m}$.

However, the true delays generally do not lie exactly on the predefined
discrete grid $\mathcal{D}$ in practice, which leads to the energy
leakage \cite{FDD2D}. To handle this issue, the authors in \cite{nsdi,CS2019,CS2020}
employ a dense grid ($L\gg N$) to make the equation (\ref{eq:ULAsignal})
hold approximately, which leads to a high computational complexity.
To overcome the energy leakage caused by delay mismatch and high computational
complexity issues of using fixed grids, we adopt a dynamic grid adjustment
strategy. Specifically, we introduce an off-grid vector $\Delta\mathbf{\boldsymbol{\tau}}=\left[\Delta\tau_{1},\cdots,\Delta\tau_{L}\right]$,
which satisfies $\Delta\tau_{l_{k}}=\tau_{k}-\overline{d}_{l_{k}},k=1,\cdots,K$
and $\Delta\tau_{l}=0,\forall l\notin\left\{ l_{1},\cdots,l_{K}\right\} $.
Note that $l_{k}\triangleq\underset{l}{\mathrm{argmin}}\left|\tau_{k}-\overline{d}_{l}\right|$
denotes the index of grid which is nearest to $\tau_{k}$. Then, $\mathbf{A}_{m}$
can be rewritten as
\begin{equation}
\mathbf{A}_{m}\left(\Delta\mathbf{\boldsymbol{\tau}}\right)=\left[\mathbf{a}_{m}\left(\overline{d}_{1}+\Delta\tau_{1}\right),\mathbf{a}_{m}\left(\overline{d}_{2}+\Delta\tau_{2}\right),\cdots,\mathbf{a}_{m}\left(\overline{d}_{L}+\Delta\tau_{L}\right)\right].
\end{equation}
Now, the mismatch can be compensated by the off-grid vector $\Delta\boldsymbol{\tau}$
and signal model (\ref{eq:ULAsignal}) can be held even if the true
delays do not lie on the uniform grid $\mathcal{D}$.

After the sparse representation, we propose a support-based probability
model to capture the structured sparsity of the multiband channels.
Since all sparse vectors $\mathbf{x}_{m}$'s share a common support
corresponding to the true delays, the sparse vector $\mathbf{x}=\left[\mathbf{x}_{1};\ldots;\mathbf{x}_{M}\right]$
obeys a group sparsity. We use $\mathbf{s}=\left[s_{1},\ldots,s_{L}\right]^{T}\in\left\{ 0,1\right\} ^{L}$
to denote the common support vector of $\mathbf{x}_{m},\forall m$,
where $s_{l}=1$ indicates that $x_{m,l},\forall m$ are active (non-zero),
while $s_{l}=0$ indicates that $x_{m,l},\forall m$ are inactive
(zero). Then, we further model the channel prior information using
a Bernoulli-Gaussian (BG) model \cite{BG,BG2}. Given the channel
support vector, $\mathbf{s}$, the conditional prior distribution
of the elements of $\mathbf{x}_{m},\forall m$ is independent and
can be written as
\begin{equation}
p\left(x_{m,l}\mid s_{l}\right)=\left(1-s_{l}\right)\delta\left(x_{m,l}\right)+s_{l}\mathcal{CN}\left(x_{m,l};0,\sigma_{m,l}^{2}\right).
\end{equation}

\subsection{Turbo-BI Algorithm}

To achieve the goal of delay estimation, we need to estimate the sparse
vector $\mathbf{x}$, support vector $\mathbf{s}$, and the uncertain
parameters $\mathbf{\boldsymbol{\xi}}\triangleq[\boldsymbol{\delta}^{T},(\Delta\boldsymbol{\mathbf{\tau}})^{T}]^{T}$,
where $\boldsymbol{\delta}=\left[\delta_{1},\ldots,\delta_{M}\right]^{T}$,
given the observations $\mathbf{y}=\left[\mathbf{y}_{1};\ldots;\mathbf{y}_{M}\right]\in\mathbb{C}^{N\times1}$.
In particular, for given $\boldsymbol{\xi}$, we are interested in
computing the conditional marginal posteriors $p(\mathbf{x}|\mathbf{y},\mathbf{\boldsymbol{\xi}})$.
For the uncertain parameters $\boldsymbol{\xi}$, we adopt the maximum
a posteriori (MAP) estimation method as

\begin{equation}
\boldsymbol{\xi}^{*}=\underset{\boldsymbol{\xi}}{\mathrm{argmax}}\thinspace p(\boldsymbol{\xi}|\mathbf{y})\propto\underset{\boldsymbol{\xi}}{\mathrm{argmax}}\thinspace\ln p(\mathbf{y},\boldsymbol{\xi})=\underset{\boldsymbol{\xi}}{\mathrm{argmax}}\thinspace\ln\underset{\mathbf{s}}{\sum}\int p(\mathbf{y},\boldsymbol{\xi},\mathbf{x},\mathbf{s})d\mathbf{x}.\label{eq:MstepMAP}
\end{equation}

This is a high-dimensional non-convex objective function and we cannot
obtain a closed-form expression due to the multi-dimensional integration
over $\mathbf{x}$ and $\mathbf{s}$, which makes it difficult to
directly maximize $\ln p(\mathbf{y},\boldsymbol{\xi})$. To handle
this issue, we adopt the majorization-minimization (MM) method to
construct a surrogate function and then use alternating optimization
(AO) method to find a stationary point of (\ref{eq:MstepMAP}). Inspired
by the expectation-maximization (EM) method \cite{2003EM}, we propose
a Turbo-BI algorithm which performs iterations between the following
two steps until convergence.
\begin{itemize}
\item Turbo-BI-E Step: For given $\mathbf{\boldsymbol{\xi}}$, approximately
calculate the posterior $p(\mathbf{x}|\mathbf{y},\mathbf{\boldsymbol{\xi}})$
by combining the message passing and linear minimum mean square error
(LMMSE) approaches via a turbo framework, as will be elaborated in
Subsection \ref{subsec:Turbo-BI-E-Step};
\item Turbo-BI-M Step: Given $p(\mathbf{x}|\mathbf{y},\mathbf{\boldsymbol{\xi}})$,
construct a surrogate function for $\ln p(\mathbf{y},\boldsymbol{\xi})$
based on the MM method, partition $\boldsymbol{\xi}$ into $B$ blocks
$\mathbf{\boldsymbol{\xi}}=(\mathbf{\boldsymbol{\xi}}_{1},...,\boldsymbol{\xi}_{B})$,
then alternatively maximize the surrogate function with respect to
$\mathbf{\boldsymbol{\xi}_{\mathit{j}}},j=1,\ldots,B$, as will be
elaborated in Subsection \ref{subsec:Turbo-BI-M-Step}.
\end{itemize}

\subsubsection{Turbo-BI-E Step\label{subsec:Turbo-BI-E-Step}}

\begin{figure}[t]
\centering{}\includegraphics[width=10cm]{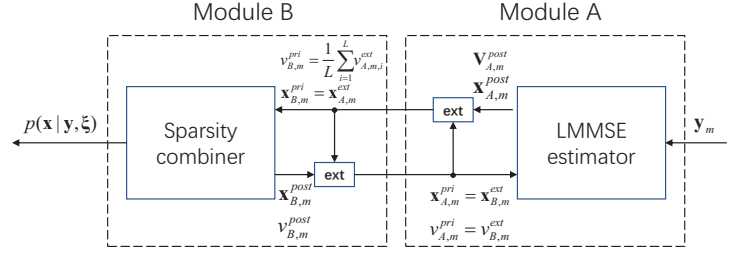}\caption{\label{fig:Turbo}Modules of the Turbo-BI-E step and message flow
between different modules.}
\end{figure}

The Turbo-BI-E step contains two modules, as illustrated in Fig. \ref{fig:Turbo}.
Module A is a LMMSE estimator based on the observation $\mathbf{y}_{m},\forall m$
and Module B is a sparsity combiner that utilizes the sparsity information
of $\mathbf{x}_{m},\forall m$ to further refine the estimation results.
The extrinsic estimation of one module will be treated as a prior
mean for the other module in the next iteration. Based on the iterations
between these two modules, the channel prior information and observations
information are combined to be exploited. Specifically, in Module
A, we assume that $\mathbf{x}_{m},\forall m$ follows a Gaussian distribution
$\mathcal{CN}\left(\mathbf{x}_{m};\mathbf{x}_{A,m}^{pri},v_{A,m}^{pri}\mathbf{I}\right)$,
where $\mathbf{x}_{A,m}^{pri}$ and $v_{A,m}^{pri}$ are the extrinsic
message output from Module B. We define $\boldsymbol{\Phi}_{m}=\mathbf{S}_{m}\mathbf{A}_{m}\left(\Delta\mathbf{\boldsymbol{\tau}}\right)\in\mathbb{C}^{N_{m}\times L}$
as the measurement matrix and we can obtain the conditional distribution
$p\left(\mathbf{y}_{m}|\mathbf{x}_{m}\right)=\mathcal{CN}\left(\boldsymbol{\Phi}_{m}\mathbf{x}_{m},\sigma_{ns}^{2}\mathbf{I}\right)$.
Then, the posterior distribution of $\mathbf{x}_{m}$ is given by
$p\left(\mathbf{x}_{m}\mid\mathbf{y}_{m}\right)=\mathcal{C}\mathcal{N}\left(\mathbf{x}_{A,m}^{post},\mathbf{V}_{A,m}^{post}\right)$,
where
\begin{equation}
\mathbf{V}_{A,m}^{post}=\left(\frac{\boldsymbol{\Phi}_{\mathit{m}}^{\mathit{H}}\boldsymbol{\Phi}_{m}}{\sigma_{ns}^{2}}+\frac{1}{v_{A,m}^{pri}}\mathbf{I}\right)^{-1},\label{eq:VApost_inv}
\end{equation}

\begin{equation}
\mathbf{x}_{A,m}^{post}=\mathbf{V}_{A,m}^{post}\left(\frac{\mathbf{x}_{A,m}^{pri}}{v_{A,m}^{pri}}+\frac{\boldsymbol{\Phi}_{m}^{H}\mathbf{y}_{m}}{\sigma_{ns}^{2}}\right).\label{eq:XAPOST}
\end{equation}

Note that in \cite{Turbo_BI}, the singular value decomposition (SVD)
of $\boldsymbol{\Phi}_{m}$ is utilized to reduce the computational
complexity of $\mathbf{V}_{A,m}^{post}$, which involves a matrix
inverse operation. In our case, however, the complexity of the matrix
inverse operation is acceptable. It is because $L\ll N_{m}$, the
complexity of the matrix inversion operation, $\mathcal{O}\left(L^{3}\right)$,
is relatively low. Finally, we calculate the extrinsic message passing:
\begin{equation}
v_{A,m,i}^{ext}=\left(\frac{1}{v_{A,m,i}^{post}}-\frac{1}{v_{A,m}^{pri}}\right)^{-1},\label{eq:vAext}
\end{equation}

\begin{equation}
x_{A,m,i}^{ext}=v_{A,m,i}^{ext}\left(\frac{x_{A,m,i}^{post}}{v_{A,m,i}^{post}}-\frac{x_{A,m,i}^{pri}}{v_{A,m}^{pri}}\right),\label{eq:xAext}
\end{equation}
where $v_{A,m,i}^{post}$ is the $i$-th diagonal element of $\mathbf{V}_{A,m}^{post}$. 

\begin{figure}[t]
\centering{}\includegraphics[width=10cm]{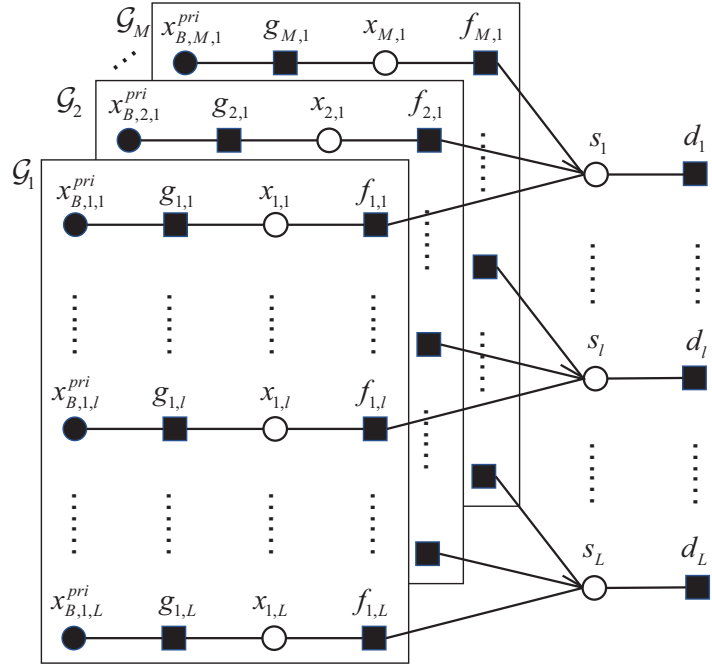}\caption{\label{fig:Factor-graph-of}Factor graph of the Turbo-BI algorithm.}
\end{figure}

In Module B, we assume that $\mathbf{x}_{B,m}^{pri}$ is modeled as
an AWGN observation of $\mathbf{x}_{m}$ \cite{modelB1,modelB2}:
\begin{equation}
\mathbf{x}_{B,m}^{pri}=\mathbf{x}_{m}+\mathbf{z}_{m},\label{eq:ModuleB-1}
\end{equation}
where $\mathbf{z}_{m}\sim\mathcal{C}\mathcal{N}\left(0,v_{B,m}^{pri}\mathbf{I}\right)$
is independent of $\mathbf{x}_{m}$, $\mathbf{x}_{B,m}^{pri}=\mathbf{x}_{A,m}^{ext}$
and $v_{B,m}^{pri}=\frac{1}{L}\sum_{i=1}^{L}v_{A,m,i}^{ext}$ are
the extrinsic message from Module A. Based on (\ref{eq:ModuleB-1}),
we combine the sparsity prior information of $\mathbf{x}_{m}$ and
the extrinsic messages from Module A, aiming at calculating the posterior
distributions $p\left(x_{m,l}\mid\mathbf{x}_{B}^{pri}\right)$ by
performing the sum-product message passing (SPMP) \cite{SPMP} over
the factor graph, where $\mathbf{x}_{B}^{pri}=[(\mathbf{x}_{B,1}^{pri})^{T},\ldots,(\mathbf{x}_{B,M}^{pri})^{T}]^{T}$.
Particularly, the factor graph of the joint distribution $p\left(\mathbf{x}_{B}^{pri},\mathbf{x},\mathbf{s}\right)$
is shown in Fig. \ref{fig:Factor-graph-of}, where the function expression
of each factor node is listed in Table \ref{tab:Factors,-distributions-and}.
At subband $m$, the factor graph is denoted by $\mathcal{G_{\mathit{m}}}$.
As can be seen, factor graphs $\mathcal{G_{\mathit{m}}}$'s share
the common support vector $\mathbf{s}$.

\begin{table*}[t]
\begin{centering}
\caption{\label{tab:Factors,-distributions-and}Factors, distributions and
functional forms in Fig. \ref{fig:Factor-graph-of}.}
\par\end{centering}
\centering{}%
\begin{tabular}{|c|c|c|}
\hline 
{\small{}Factor} & {\small{}Distribution} & {\small{}Functional form}\tabularnewline
\hline 
\hline 
$g_{m,l}\left(x_{B,m,l}^{pri},x_{m,l}\right)$ & $p\left(x_{B,m,l}^{pri}\mid x_{m,l}\right)$ & $\mathcal{CN}\left(x_{m,l};x_{B,m,l}^{pri},v_{B,m}^{pri}\right)$\tabularnewline
\hline 
$f_{m,l}\left(s_{l},x_{m,l}\right)$ & $p\left(x_{m,l}\mid s_{l}\right)$ & $\left(1-s_{l}\right)\delta\left(x_{m,l}\right)+s_{l}\mathcal{CN}\left(x_{m,l};0,\sigma_{m,l}^{2}\right)$\tabularnewline
\hline 
$d_{l}\left(s_{l}\right)$ & $p\left(s_{l}\right)$ & $p_{s}$\tabularnewline
\hline 
\end{tabular}
\end{table*}

We now outline the message passing scheme on graph $\mathcal{G}$.
The details are elaborated in Appendix \ref{subsec:A.-Message-Passing}.
According to the sum-product rule, the message passing over the path
$x_{m,l}\rightarrow f_{m,l}\rightarrow s_{l}$ are given by (\ref{eq:MP1})
and (\ref{eq:MP2}). Then the message is passed back over the path
$s_{l}\rightarrow f_{m,l}\rightarrow x_{m,l}$ using (\ref{eq:MP3})
and (\ref{eq:MP4}). After calculating the updated messages $\left\{ v_{f_{m,l}\rightarrow x_{m,l}}\right\} $,
the approximate posterior distributions are given by
\begin{equation}
p\left(x_{m,l}\mid\mathbf{x}_{B}^{pri}\right)\wasypropto v_{f_{m,l}\rightarrow x_{m,l}}\left(x_{m,l}\right)v_{g_{m,l}\rightarrow x_{m,l}}\left(x_{m,l}\right),\label{eq:MP5}
\end{equation}
where $v_{g_{m,l}\rightarrow x_{m,l}}\left(x_{m,l}\right)=\mathcal{C}\mathcal{N}\left(x_{m,l};x_{B,m,l}^{pri},v_{B,m}^{pri}\right)$.
Then the posterior mean and variance are given by
\begin{equation}
x_{B,m,l}^{post}=\mathbb{E}\left(x_{m,l}\mid\mathbf{x}_{B}^{pri}\right)=\int x_{m,l}\thinspace p\left(x_{m,l}\mid\mathbf{x}_{B}^{pri}\right)dx_{m,l},\label{eq:MP6}
\end{equation}

\begin{equation}
v_{B,m}^{post}=\frac{1}{L}\sum_{l=1}^{L}\mathrm{Var}\left(x_{m,l}\mid\mathbf{x}_{B}^{pri}\right)=\frac{1}{L}\sum_{l=1}^{L}\int\left|x_{m,l}-\mathbb{E}\left(x_{m,l}\mid\mathbf{x}_{B}^{pri}\right)\right|^{2}p\left(x_{m,l}\mid\mathbf{x}_{B}^{pri}\right)dx_{m,l}.\label{eq:MP7}
\end{equation}
Finally, based on the derivation in \cite{XApri}, the extrinsic update
for Module A can be calculated as
\begin{equation}
\mathbf{x}_{A,m}^{pri}=\mathbf{x}_{B,m}^{ext}=v_{A,m}^{pri}\left(\frac{\mathbf{x}_{B,m}^{post}}{v_{B,m}^{post}}-\frac{\mathbf{x}_{B,m}^{pri}}{v_{B,m}^{pri}}\right),\label{eq:xBext}
\end{equation}

\begin{equation}
v_{A,m}^{pri}=v_{B,m}^{ext}=\left(\frac{1}{v_{B,m}^{post}}-\frac{1}{v_{B,m}^{pri}}\right)^{-1}.\label{eq:vBext}
\end{equation}

\subsubsection{Turbo-BI-M Step\label{subsec:Turbo-BI-M-Step}}

In the M-step, we construct a surrogate function at fixed point $\dot{\boldsymbol{\xi}}$
for the objective function (\ref{eq:MstepMAP}) based on the MM method
as
\begin{equation}
u(\mathbf{\boldsymbol{\xi}};\dot{\mathbf{\boldsymbol{\xi}}})=\int p(\mathbf{x}\mid\mathbf{y},\dot{\mathbf{\xi}})\ln\frac{p(\mathbf{x},\mathbf{y},\mathbf{\boldsymbol{\xi}})}{p(\mathbf{x}\mid\mathbf{y},\dot{\mathbf{\boldsymbol{\xi}}})}d\mathbf{x},\label{eq:surrogatefunc}
\end{equation}
which satisfies basic properties $u(\boldsymbol{\mathbf{\xi}};\dot{\boldsymbol{\mathbf{\xi}}})\leq\ln p(\mathbf{y},\boldsymbol{\mathbf{\xi}}),\forall\boldsymbol{\mathbf{\xi}}$;
$u(\dot{\mathbf{\boldsymbol{\xi}}};\dot{\mathbf{\boldsymbol{\xi}}})=\ln p(\mathbf{y},\dot{\mathbf{\xi}}),\forall\boldsymbol{\mathbf{\xi}}$;
$\left.\frac{\partial u(\mathbf{\boldsymbol{\mathbf{\xi}}};\dot{\mathbf{\boldsymbol{\xi}}})}{\partial\boldsymbol{\xi}}\right|_{\mathbf{\boldsymbol{\xi}}=\dot{\mathbf{\boldsymbol{\xi}}}}=\left.\frac{\partial\ln p(\boldsymbol{y},\mathbf{\boldsymbol{\xi}})}{\partial\boldsymbol{\xi}}\right|_{\boldsymbol{\mathbf{\xi}}=\dot{\mathbf{\boldsymbol{\xi}}}},\forall\boldsymbol{\mathbf{\xi}}$.
Then, we partition $\mathbf{\boldsymbol{\xi}}$ into $B=2$ blocks
with $\boldsymbol{\xi}_{1}=\boldsymbol{\delta}$, $\boldsymbol{\xi}_{2}=\Delta\boldsymbol{\tau}$
based on their distinct physical meaning, and alternatively update
$\boldsymbol{\xi}_{1}$ and $\boldsymbol{\xi}_{2}$ as
\begin{eqnarray}
\mathbf{\boldsymbol{\delta}}^{(i+1)} & = & \underset{\mathbf{\boldsymbol{\delta}}}{\mathrm{argmax}}\thinspace u\left(\mathbf{\boldsymbol{\delta}},\Delta\boldsymbol{\tau}^{(i)};\mathbf{\boldsymbol{\delta}}^{(i)},\Delta\boldsymbol{\tau}^{(i)}\right),\label{eq:Mstep_delta}\\
\Delta\mathbf{\boldsymbol{\tau}}^{(i+1)} & = & \underset{\boldsymbol{\Delta}\mathbf{\tau}}{\mathrm{argmax}}\thinspace u\left(\boldsymbol{\delta}^{(i+1)},\Delta\boldsymbol{\tau};\boldsymbol{\delta}^{(i)},\Delta\boldsymbol{\tau}^{(i)}\right).\label{Mstep_tau}
\end{eqnarray}

Since the optimization problems (\ref{eq:Mstep_delta}) and (\ref{Mstep_tau})
are non-convex and it is hard to find their optimal solutions, we
apply a one-step gradient update for $\mathbf{\boldsymbol{\delta}}$
and $\Delta\boldsymbol{\mathbf{\tau}}$ as follows:
\begin{eqnarray}
\boldsymbol{\delta}^{(i+1)} & = & \boldsymbol{\delta}^{(i)}+\gamma_{\boldsymbol{\delta}}\cdot\boldsymbol{\zeta}_{\boldsymbol{\delta}}^{(i)},\label{eq:updaterule_delta}\\
\Delta\boldsymbol{\tau}^{(i+1)} & = & \Delta\boldsymbol{\tau}^{(i)}+\gamma_{\Delta\boldsymbol{\tau}}\cdot\boldsymbol{\zeta}_{\Delta\boldsymbol{\tau}}^{(i)},\label{eq:updaterule_tau}
\end{eqnarray}
where $\gamma_{\boldsymbol{\delta}}$ and $\gamma_{\Delta\boldsymbol{\tau}}$
are the step size determined by the Armijo rule \cite{Bertsekas_book95_NProgramming},
$\boldsymbol{\zeta}_{\boldsymbol{\delta}}^{(i)}$ and $\boldsymbol{\zeta}_{\Delta\boldsymbol{\tau}}^{(i)}$
are the gradients of the objective function in (\ref{eq:Mstep_delta})
and (\ref{Mstep_tau}) with respect to $\boldsymbol{\delta}$ and
$\Delta\boldsymbol{\tau}$, respectively. The detailed derivations
for $\boldsymbol{\zeta}_{\boldsymbol{\delta}}^{(i)}$ and $\boldsymbol{\zeta}_{\Delta\boldsymbol{\tau}}^{(i)}$
are presented in Appendix \ref{subsec:B.-Gradient-derivation}. Moreover,
the convergence of this in-exact MM algorithm to a stationary point
can be guaranteed \cite[Theorem 1]{robust_recovery}.

\subsection{Summary of the Turbo-BI Algorithm and Complexity Analysis}

The Turbo-BI algorithm is summarized in Algorithm \ref{alg:Turbo-BI-algorithm}.
Finally, we analyze the computational complexity of the proposed Turbo-BI
algorithm. It is observed that the computational complexity of the
Turbo-BI-E step is dominated by the matrix multiplication $\boldsymbol{\Phi}_{\mathit{m}}^{\mathit{H}}\boldsymbol{\Phi}_{m}$
in (\ref{eq:VApost_inv}), which is $\mathcal{O}\left(N_{m}L^{2}\right)$.

In the Turbo-BI-M step, the computational complexity of choosing the
right step size mainly depends on calculating the cost function. We
denote the number of calculating the cost function for $\gamma_{\boldsymbol{\delta}}$
and $\gamma_{\Delta\boldsymbol{\tau}}$ in every backtracking line
search as $R_{b,1}$ and $R_{b,2}$, respectively. Then, the complexity
of choosing $\gamma_{\boldsymbol{\delta}}$ and $\gamma_{\Delta\boldsymbol{\tau}}$
are $\mathcal{O}\left(NL^{2}R_{b,1}\right)$ and $\mathcal{O}\left(NL^{2}R_{b,2}\right)$,
respectively. Besides, the complexity in calculating $\boldsymbol{\zeta}_{\boldsymbol{\delta}}$
and $\boldsymbol{\zeta}_{\Delta\boldsymbol{\tau}}$ are $\mathcal{O}\left(NL^{2}\right)$
and $\mathcal{O}\left(NL\right)$ based on matrix multiplication,
respectively. Hence, the overall computational complexity of Turbo-BI
algorithm is $\mathcal{O}\left(NL^{2}I_{in}+NL^{2}(R_{b,1}+R_{b,2})\right)$
per iteration, where $I_{in}$ denotes the number of Turbo iterations
for convergence.
\begin{algorithm}[t]
{\small{}\caption{\label{alg:Turbo-BI-algorithm}Turbo-BI algorithm}
}{\small\par}

\textbf{Input:} $\boldsymbol{y}$, $\boldsymbol{\Phi}_{m},\forall m$,
\textcolor{black}{maximum iteration number $I_{EM}$, threshold $\epsilon$.}

\textbf{Output:} $\boldsymbol{\delta}^{\ast}$, $\Delta\boldsymbol{\tau}^{\ast}$.

\begin{algorithmic}[1]

\FOR{${\color{blue}{\color{black}i=1,\cdots,I_{EM}}}$}

\STATE \textbf{Turbo-BI-E Step:}

\WHILE{not converge}

\STATE \textbf{\%Module A: LMMSE Estimator}

\STATE Initialize $\mathbf{x}_{A,m}^{pri}=\boldsymbol{0}$ and $v_{A,m}^{pri}$.

\STATE Update $\mathbf{V}_{A,m}^{post}$ and $\mathbf{x}_{A,m}^{post}$,
using (\ref{eq:VApost_inv}) and (\ref{eq:XAPOST}).

\STATE Update the extrinsic information $v_{B,m}^{pri}=\frac{1}{L}\sum_{i=1}^{L}v_{A,m,i}^{ext}$
and $\mathbf{x}_{B,m}^{pri}=\mathbf{x}_{A,m}^{ext}$, using (\ref{eq:vAext})
and (\ref{eq:xAext}).

\STATE \textbf{\% Module B: Sparsity Combiner}

\STATE Perform message passing over the factor graph $\mathcal{G}$
using (\ref{eq:MP1}) - (\ref{eq:MP4}).

\STATE Calculate the approximate posterior distributions $p\left(x_{l}^{m}\mid\mathbf{x}_{B}^{pri}\right)$
using (\ref{eq:MP5}).

\STATE Update $x_{B,m,l}^{post}$ and $\ensuremath{v_{B,m}^{post}}$
using (\ref{eq:MP6}) and (\ref{eq:MP7}).

\STATE Update the extrinsic information $\mathbf{x}_{A,m}^{pri}=\mathbf{x}_{B,m}^{ext}$
and $v_{A,m}^{pri}=v_{B,m}^{ext}$, using (\ref{eq:xBext}) and (\ref{eq:vBext}).

\ENDWHILE

\STATE \textbf{Turbo-BI-M Step:}

\STATE Construct the surrogate function in (\ref{eq:surrogatefunc})
using $\mathbf{x}_{A,m}^{post}$ and $\mathbf{V}_{A,m}^{post}$, which
is from the Turbo-BI-E step.

\STATE Update $\boldsymbol{\delta}$ and $\Delta\boldsymbol{\tau}$,
using (\ref{eq:updaterule_delta}) and (\ref{eq:updaterule_tau}).

\IF{\textcolor{black}{{} }$\left\Vert \Delta\boldsymbol{\delta}^{(i+1)}-\Delta\boldsymbol{\delta}^{(i)}\right\Vert \leq\epsilon$
and $\left\Vert \Delta\boldsymbol{\tau}^{(i+1)}-\Delta\boldsymbol{\tau}^{(i)}\right\Vert \leq\epsilon$}

\STATE \textbf{\textcolor{black}{break}}

\ENDIF

\ENDFOR

\end{algorithmic}
\end{algorithm}

\section{\label{sec:PSO-LS-algorithm-in}PSO-LS Algorithm in Stage 2}

In Stage 2, we aim to fully exploit the frequency band apertures gain
and perform a refined delay estimation based on the estimation result
from Stage 1. First, we reformulate the refined estimation signal
model (\ref{eq:refined_signal}) as a linear form:
\begin{equation}
\mathbf{y}=\mathbf{H\left(\boldsymbol{\theta}\right)x}+\boldsymbol{w},
\end{equation}
where $\mathbf{H\left(\boldsymbol{\theta}\right)}=\left[\begin{array}{ccc}
\mathbf{h}_{11} & \cdots & \mathbf{h}_{1K}\\
\vdots & \ddots & \vdots\\
\mathbf{h}_{M1} & \cdots & \mathbf{h}_{MK}
\end{array}\right]\in\mathbb{C}^{N\times K}$, $h_{mk}(n)=e^{-j2\pi\left(f_{c,m}^{\prime}+nf_{s,m}\right)\tau_{k}}e^{-j2\pi nf_{s,m}\delta_{m}}e^{j\varphi_{m}^{\prime}}$
denotes the $n$-th element of the column vector $\mathbf{h}_{mk}$,
$\mathbf{\boldsymbol{\theta}}=[\tau_{1},...,\tau_{K},$$\delta_{1},...,\delta_{M},\varphi_{2}^{\prime},...,\varphi_{M}^{\prime}]^{T}\in\mathbb{R}^{(K+2M-1)\times1}$
denotes the vector consisting of unknown parameters, $\mathbf{x}=\left[\alpha_{1}^{\prime},\ldots,\alpha_{K}^{\prime}\right]^{T}\in\mathbb{C}^{K\times1}$,
and $\boldsymbol{w}\sim\mathcal{C}\mathcal{N}(0,\sigma_{ns}^{2}\mathbf{I})\in\mathbb{C}^{N\times1}$.
We adopt the MAP method for estimation, that takes the prior information
of $\delta_{m}$ into consideration. Then, the optimization problem
can be formulated as
\begin{eqnarray}
\mathcal{P}_{1}: & \underset{\boldsymbol{\theta},\mathbf{x}}{\max} & \ln\thinspace p\left(\mathbf{y}|\mathbf{x}\right)+\sum_{m=1}^{M}\textrm{ln}\thinspace p\left(\delta_{m}\right)\nonumber \\
 & \text{s.t. } & 0\leq\varphi_{m}^{\prime}\leq2\pi,\forall m\in\left\{ 2,\cdots,M\right\} ,\label{eq:P1}
\end{eqnarray}
where $p\left(\mathbf{y}|\mathbf{x}\right)\wasypropto e^{-\frac{\|\mathbf{y}-\mathbf{H\left(\theta\right)x}\|^{2}}{\sigma_{ns}^{2}}}$
is the likelihood function. After an equivalent transformation, $\mathcal{P}_{1}$
can be reformulated as
\begin{eqnarray}
\mathcal{P}_{2}: & \underset{\boldsymbol{\theta},\mathbf{x}}{\min} & \frac{\|\mathbf{y}-\mathbf{H\left(\boldsymbol{\theta}\right)x}\|^{2}}{\sigma_{ns}^{2}}+\sum_{m=1}^{M}\frac{\delta_{m}^{2}}{2\sigma_{p}^{2}}\nonumber \\
 & \text{\textrm{s.t.}} & 0\leq\varphi_{m}^{\prime}\leq2\pi,\forall m\in\left\{ 2,\cdots,M\right\} .\label{eq:P2}
\end{eqnarray}

The non-convex problem $\mathcal{P}_{2}$ has a multimodal and multi-dimensional
objective function, which is extremely challenging to solve due to
the existence of numerous local optimums. In this case, conventional
algorithms, such as gradient descent and exhaustive search algorithms,
have high-computational complexity or are easily trapped into local
optimums. To overcome these drawbacks, we adopt the PSO method \cite{PSO1,PSO2,PSOCL3,PSOHeterogeneous4}
to find a good solution for the non-convex optimization problem $\mathcal{P}_{2}$.

In the PSO method, we employ a number of particles which are potential
solutions, to find the optimal solution by iterations, where the search
space is bounded by the constraints of the target optimization problem
\cite{2001Swarm}. This heuristic algorithm has been shown to have
low computational complexity and a strong optimization ability for
complex multimodal optimization problems \cite{PSOCL3}. Generally,
PSO starts with a random initialization of the particles' locations
in a large search space. In our problem $\mathcal{P}_{2}$, however,
we can narrow down the search space based on the coarse estimation
results from Stage 1.

Specifically, we set the search space as 
\begin{equation}
\mathcal{S}=[\boldsymbol{\beta}-\mathbf{e},\boldsymbol{\beta}+\mathbf{e}]\in\mathbb{R}^{D\times1},\label{eq:searchspace}
\end{equation}
where $\boldsymbol{\beta}$ has the values consisting of the coarse
estimation results, $\mathbf{e}$ is the search range obtained by
evaluating the mean squared error (MSE) of coarse estimation results
based on offline training or evaluating the Cram�r-Rao bound (CRB)
values based on coarse signal model (\ref{eq:coarse_signal}), and
$D$ denotes the dimension of the particles (search space). Since
there is no estimation for $\varphi_{m}^{\prime}$ in Stage 1, the
search space for $\varphi_{m}^{\prime}$ is set to be $\left[0,2\pi\right]$.
Note that the search space $S$ is a real set with dimension $D=3K+2M-1$,
since the $K$-dimension complex vector $\mathbf{x}$ can be seen
as a $2K$-dimension real vector in a real domain search space. Then,
for the $i$-th iteration, each particle comes to a better location
by updating its velocity and position based on the following two equations:
\begin{equation}
V_{q,d}^{(i+1)}=\omega V_{q,d}^{(i)}+c_{1}r_{1,q,d}\left(pbest_{q,d}^{(i)}-X_{q,d}^{(i)}\right)+c_{2}r_{2,q,d}\left(gbest_{d}^{(i)}-X_{q,d}^{(i)}\right),\label{eq:velocityupdate}
\end{equation}

\begin{equation}
X_{q,d}^{(i+1)}=X_{q,d}^{(i)}+V_{q,d}^{(i+1)},\label{eq:positionupdate}
\end{equation}
where $c_{1}$ and $c_{2}$ are acceleration coefficients, and $\omega$
is the inertia factor, which is proposed to balance the global and
local search abilities \cite{PSO_first}. Note that $X_{q,d}^{(i)}$
and $V_{q,d}^{(i)}$ are the position and velocity of the $d$-th
element of the $q$-th particle, respectively, where $q\in\left\{ 1,2,...,Q_{p}\right\} ,d\in\left\{ 1,2,...,D\right\} $
with $Q_{p}$ denoting the number of particles. $r_{1,q,d}$ and $r_{2,q,d}$
are two independent random numbers uniformly distributed within $[0,1]$.
Besides, $gbest$ denotes the best position in the whole swarm and
$pbest_{q,d}^{(i)}$ denotes the best position values at the $d$-th
dimension that the $q$-th particle has come across so far at the
$i$-th iteration. The goodness of the particle position is measured
by the objective function value in $\mathcal{P}_{2}$ (also called
fitness in PSO algorithms). In our problem, a smaller fitness indicates
a better position.

For getting a better solution in the PSO algorithm, we aim to reduce
the dimension of the search space in problem $\mathcal{P}_{2}$ and
thus propose the PSO-LS algorithm. We first employ primal-decomposition
theory to decouple the problem $\mathcal{P}_{2}$ as:
\begin{eqnarray}
\mathcal{P}_{3}: & \underset{\boldsymbol{\theta}}{\min} & \frac{\|\mathbf{y}-\mathbf{H\left(\boldsymbol{\theta}\right)\mathit{g}^{*}(\boldsymbol{\theta})}\|^{2}}{\sigma_{ns}^{2}}+\sum_{m=1}^{M}\frac{\delta_{m}^{2}}{2\sigma_{p}^{2}}\nonumber \\
 & \text{s.t. } & 0\leq\varphi_{m}^{\prime}\leq2\pi,\forall m\in\left\{ 2,\cdots,M\right\} ,\label{eq:LSPSO}
\end{eqnarray}
where $g^{*}(\boldsymbol{\theta})=\underset{\mathbf{x}}{\mathrm{argmin}}\|\mathbf{y}-\mathbf{H\left(\boldsymbol{\theta}\right)x}\|^{2}$.
Then we can obtain a least-square (LS) solution of $g^{*}(\boldsymbol{\theta})$
as
\begin{equation}
g^{*}(\boldsymbol{\theta})=\left(\mathbf{H\left(\boldsymbol{\theta}\right)}^{H}\mathbf{H\left(\boldsymbol{\theta}\right)}\right)^{-1}\mathbf{H\left(\boldsymbol{\theta}\right)}^{H}\mathbf{y}.
\end{equation}
Using this method, the dimension of search space is reduced from $\mathbb{R}^{3K+2M-1}$
to $\mathbb{R}^{K+2M-1}$. Then, we analyze the computational complexity
of the PSO-LS algorithm. The complexity mainly depends on the calculations
of the objective function in (\ref{eq:LSPSO}), which is $\mathcal{O}\left(NK^{2}Q_{p}\right)$
per iteration. Finally, the proposed PSO-LS algorithm is presented
in Algorithm \ref{alg:PSO-LSalgorithm}.
\begin{algorithm}[t]
{\small{}\caption{\label{alg:PSO-LSalgorithm}PSO-LS algorithm}
}{\small\par}

\textbf{Input:} $\mathbf{y}$, search space $\mathcal{S}$, $Q_{p}$,\textcolor{black}{{}
maximum iteration number $I_{PSO}$, threshold $\epsilon$.}

\textbf{Output:} $\mathbf{gbest}$.

\begin{algorithmic}[1]

\STATE Initialize $X_{q,d}^{(1)}$ within the search space $\mathcal{S}$
in (\ref{eq:searchspace}), $V_{q,d}^{(1)}$, $gbest_{d}^{(1)}$,
$pbest_{q,d}^{(1)}$, $\forall q,d$.

\FOR{ $i=1,\cdots,I_{PSO}$}

\FOR{${\color{blue}{\color{black}q=1,\cdots,Q_{p}}}$}

\STATE Update the velocity of particle $q$ using (\ref{eq:velocityupdate}).

\STATE Update the position of particle $q$ using (\ref{eq:positionupdate}).

\STATE Calculate the fitness of particle $q$ based on the objective
in (\ref{eq:LSPSO}).

\STATE Update $pbest_{q,d}^{(i+1)}$ and $gbest_{d}^{(i+1)}$ ,$\forall q,d$.

\ENDFOR

\IF{\textcolor{black}{{} }$\left\Vert \mathbf{gbest}^{(i+1)}-\mathbf{gbest}^{(i)}\right\Vert \leq\epsilon$}

\STATE \textbf{\textcolor{black}{break}}

\ENDIF

\ENDFOR

\end{algorithmic}
\end{algorithm}

\section{\label{sec:Simulation-Results}Simulation Results}

In this section, we provide numerical results to evaluate the performance
of the proposed schemes and draw useful insights. In the default setup,
we consider that CSI samples are collected using one OFDM training
symbol with subcarrier spacing $f_{s,1}=f_{s,2}=60$ KHz and subband
bandwidth $B_{1}=B_{2}=40$ MHz at $M=2$ subbands, with central frequencies
$f_{c,1}=1.80$ GHz and $f_{c,2}=2.02$ GHz. $\varphi_{m},\forall m$
and $\delta_{m},\forall m$ are generated following a uniform distribution
within $[0,2\pi]$ and a Gaussian distribution $\mathcal{N}\left(0,\sigma_{p}^{2}\right)$,
respectively. We use the Quadriga platform to generate multiband CSI
samples in an indoor factory (InF) scenario, which is depicted in
3GPP R16 \cite{3gpp_Rel16}. Other system parameters are set as follows
unless otherwise specified: \textcolor{black}{Signal-to-noise ratio
(SNR) is 7 dB}\textcolor{black}{\small{}, $I_{EM}=100$, $I_{PSO}=500$},
$Q_{p}=100$, $\epsilon=10^{-5}$, $c_{1}=2.5$, $c_{2}=0.5$, and
$\omega=0.99-\frac{0.79}{I_{PSO}}t$ for the $t$-th PSO iteration.\textcolor{black}{\small{}
}To assess the performance of the schemes, we adopt the empirical
cumulative distribution function (CDF) of the LoS path delay estimation
errors using 200 Monte Carlo trials.

For comparison, we consider the following four benchmark schemes:
\begin{itemize}
\item \textbf{Turbo-BI algorithm:} We adopt the coarse estimation results
of the Turbo-BI algorithm in Stage 1 as one of the benchmarks.
\item \textbf{Multiband weighted delay estimation (MBWDE) algorithm \cite{ESPRIT2}:}
It adopts a weighted subspace fitting algorithm for delay estimation,
which is able to exploit the frequency band apertures gain.
\item \textbf{Two-stage gradient descent (TSGD) scheme:} This scheme has
the same coarse estimation implementation as the TSGE scheme in Stage
1 and then the gradient descent method is employed to perform refined
estimation in Stage 2 with the initial point generated using the coarse
estimation in Stage 1.
\item \textbf{Ideal gradient descent (IGD) scheme:} This scheme adopts gradient
descent method based on problem $\mathcal{P}_{2}$, where we set the
initial point to be the true values. In fact, this scheme is infeasible
because we cannot know the true values of the estimation parameters
in practical environment. In the simulations, however, we can regard
this scheme as an ideal benchmark to show the effectiveness of the
TSGE scheme.
\end{itemize}

\subsection{Impact of the factor $\boldsymbol{\delta}$}

\begin{figure}[t]
\centering{}\includegraphics[width=10cm]{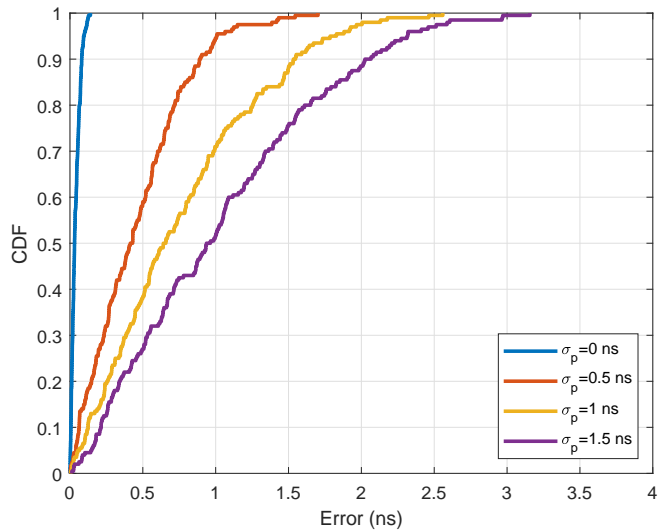}\caption{\label{fig:CDF_delta}The CDF of LoS path delay estimation errors
for different $\sigma_{p}$. }
\end{figure}

We first study the impact of the receiver timing offset $\boldsymbol{\delta}$.
Since the InF scenario does not consider the effect of $\boldsymbol{\delta}$,
we construct a two-path channel model with Rayleigh distributed magnitudes.
The delays are set to follow a uniform distribution within $[20,200]$
ns. Fig. \ref{fig:CDF_delta} depicts the CDF of LoS path delay estimation
errors for different standard deviation $\sigma_{p}$ achieved by
TSGE. As can be seen, the factor $\boldsymbol{\delta}$ has significantly
degraded the estimation performance. Besides, when $\boldsymbol{\delta}$
is considered, the estimation performance mainly depends on the prior
standard deviation $\sigma_{p}$, where the delay estimation errors
increase with $\sigma_{p}$. It is reasonable since a larger $\sigma_{p}$
means less prior information of $\boldsymbol{\delta}$ we have. Therefore,
in the following simulations, we assume that $\sigma_{p}=0$ ns in
order to avoid its effect on the estimation performance.

\subsection{Performance of TSGE Scheme}

\begin{figure}[t]
\centering{}\includegraphics[width=10cm]{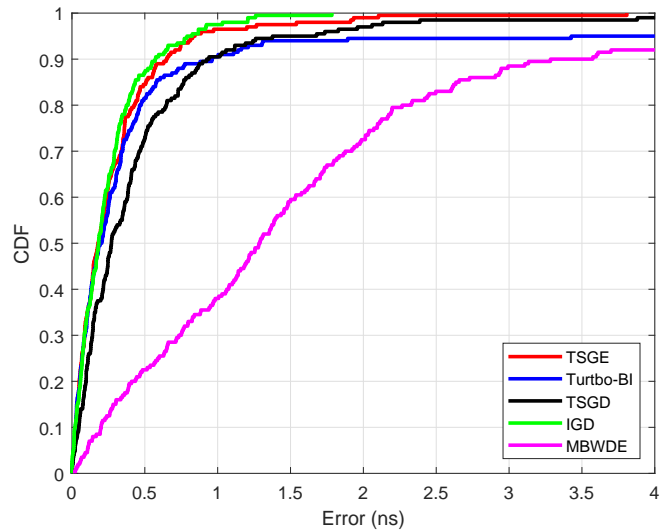}\caption{\label{fig:Performance compare}The CDF of LoS path delay estimation
errors for different schemes.}
\end{figure}

In Fig. \ref{fig:Performance compare}, we compare the delay estimation
performance of the TSGE scheme with benchmarks. First, we observe
that the CS-based schemes (i.e., TSGE, Turbo-BI, and TSGD) and the
IGD scheme achieve a better performance than the subspace-based algorithm,
i.e., MBWDE. It is mainly because that the received CSI samples are
collected using only a single OFDM training symbol, which leads to
a limited ability of subspace-based algorithms to suppress the noise
interference. In contrast, CS-based schemes have a strong ability
of reducing the effects of noise by signal sparse reconstruction,
thus they achieve a better performance. Second, it can be seen that
the proposed TSGE scheme significantly outperforms the Turbo-BI and
TSGD. This is because TSGE scheme exploits the extra frequency band
apertures gain compared to the Turbo-BI algorithm. Moreover, note
that the likelihood function based on the refined signal model (\ref{eq:refined_signal})
has numerous local optimums, which requires a strong global search
ability for the algorithms in Stage 2. Compared to the gradient descent
algorithm in the TSGD scheme, the PSO-LS algorithm in the TSGE scheme
has stronger global search ability, thus achieves higher estimation
accuracy than TSGD. Finally, it is observed that the CDF curve of
TSGE approaches that of IGD, which indicates a negligible performance
loss.

In Table \ref{tab:CPU-time-and}, we investigate the computing cost
and the estimation accuracy of the considered algorithms, i.e., the
PSO-LS algorithm with different numbers of particles and iterations,
primal PSO algorithm based on the problem $\mathcal{P}_{2}$, and
the MBWDE algorithm. The computing cost is characterized by the spent
CPU time when running the algorithms with the Intel Xeon 6248R CPU.
For fairness, we set the same initial estimation values for all algorithms.
It can be readily seen that the PSO-LS is able to achieve a higher
estimation accuracy with less computing cost than primal PSO, which
validates the effectiveness of the proposed PSO-LS algorithms. In
addition, our proposed PSO-LS algorithm is able to achieve better
estimation accuracy than MBWDE with nearly equal computing cost. Furthermore,
we can see that the PSO-LS algorithm is able to achieve a trade-off
between the computing cost and estimation performance, which implies
that it can be adapted to various scenarios. In particular, when we
have sufficient computing resource, we can use a large number of particles
and iterations to pursue the highest estimation accuracy. Conversely,
when the computing resource is limited, we can use a small number
of particles and iterations to achieve a relatively accurate estimation.

\begin{table*}[t]
\begin{centering}
\caption{\label{tab:CPU-time-and}CPU time and RMSE comparison.}
\par\end{centering}
\centering{}%
\begin{tabular}{|c|>{\centering}m{3.3cm}|>{\centering}m{3.3cm}|>{\centering}m{3.4cm}|c|}
\hline 
 & \multicolumn{1}{>{\centering}m{3.3cm}|}{{\small{}PSO-LS}{\small\par}

{\footnotesize{}(}\textcolor{black}{\footnotesize{}$I_{PSO}=500$}{\footnotesize{},$Q_{p}=100$)}} & {\small{}PSO-LS}{\small\par}

{\footnotesize{}(}\textcolor{black}{\footnotesize{}$I_{PSO}=100$}{\footnotesize{},
$Q_{p}=20$)} & {\small{}primal PSO}{\small\par}

{\footnotesize{}(}\textcolor{black}{\footnotesize{}$I_{PSO}=500$}{\footnotesize{},
$Q_{p}=100$)} & {\small{}MBWDE}\tabularnewline
\hline 
{\small{}CPU time (s)} & {\small{}32.8} & {\small{}1.3} & {\small{}20.3} & {\small{}0.8}\tabularnewline
\hline 
{\small{}RMSE (ns)} & {\small{}0.2876} & {\small{}0.5425} & {\small{}0.5517} & {\small{}1.671}\tabularnewline
\hline 
\end{tabular}
\end{table*}

\subsection{Impact of Frequency Band Spacing}

\begin{figure}[t]
\centering{}\includegraphics[width=10cm]{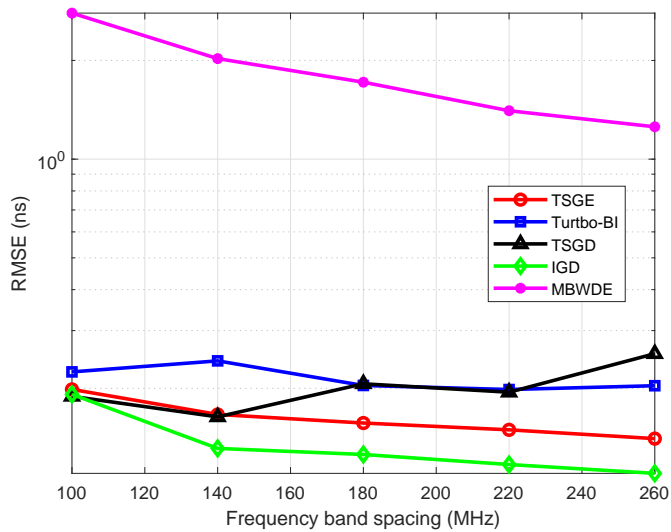}\caption{\label{fig:Impact-frequency}RMSE of LoS delay estimation versus frequency
band spacing.}
\end{figure}

Fig. \ref{fig:Impact-frequency} illustrates the root mean square
error (RMSE) of the LoS delay estimation for different schemes versus
frequency band spacing, when $\textrm{SNR}=10$ dB and bandwidth $B_{1}=B_{2}=60$
MHz. We define $\mathrm{RMSE}(\hat{\tau})\triangleq\sqrt{\mathbb{E}\left\{ (\hat{\tau}-\tau)^{2}\right\} }$,
where $\hat{\tau}$ is the estimated LoS delay and $\tau$ is the
true LoS delay. In particular, we fix $f_{c,1}$ and change $f_{c,2}$
for different frequency band spacings. It can be seen that the RMSE
decreases with the increase of frequency band spacing for MBWDE, TSGE,
and IGD. This is because these schemes are able to exploit frequency
band apertures gain, which enlarges as frequency band spacing increases.
Furthermore, the TSGD scheme has a better performance than the Turbo-BI
algorithm when frequency band spacing is narrow, but the performance
gap decreases as the frequency band spacing increases. When the frequency
band spacing is 260 MHz, the performance of TSGD is even worse than
Turbo-BI. This is reasonable since the oscillation of the likelihood
function becomes more violent with the increase of frequency band
spacing, which causes more bad local optimums and makes it more difficult
to fully exploit the frequency band apertures gain. However, TSGD
has a limited global search ability, which results in a high probability
of being stuck into local optimums. Hence, TSGD can only exploit part
of frequency band apertures gain and has a poor delay estimation performance
with large frequency band spacing. Besides, TSGE has more stable performance
than TSGD due to its strong global optimization ability. Finally,
we observe that the performance of Turbo-BI is irrelevant to the frequency
band spacing since it only exploits subcarrier apertures gain in Stage
1.

\subsection{Impact of SNR}

\begin{figure}[t]
\centering{}\includegraphics[width=10cm]{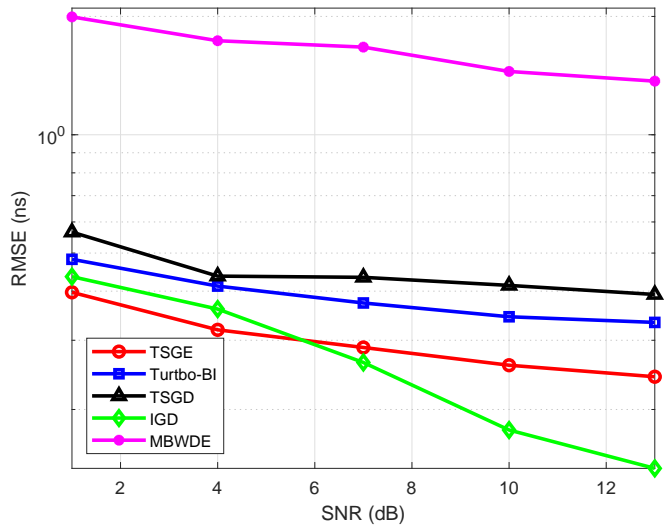}\caption{\label{fig:Impact-snr}RMSE of LoS delay estimation versus SNR.}
\end{figure}

In Fig. \ref{fig:Impact-snr}, we show the impact of SNR on the delay
estimation performance. As can be seen, the RMSE of all schemes decreases
with the increase of SNR because of the reduction of noise interference
to delay estimation. Besides, the performance gap of TSGE and IGD
over Turbo-BI becomes larger as SNR increases.

\subsection{Impact of the Bandwidth}

\begin{figure}[t]
\centering{}\includegraphics[width=10cm]{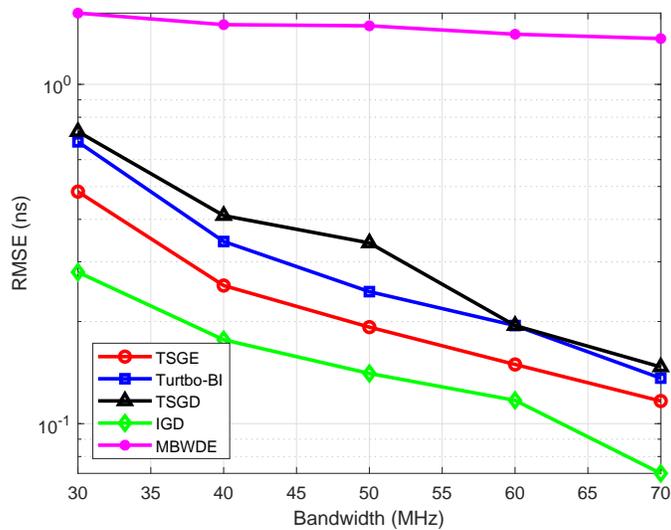}\caption{\label{fig:Impact-Bandwidth}RMSE of LoS delay estimation versus the
bandwidth.}
\end{figure}

In Fig. \ref{fig:Impact-Bandwidth}, we investigate the RMSE of the
LoS delay estimation versus the bandwidth with $\textrm{SNR}=10$
dB. It is observed that the delay estimation accuracy increases as
the bandwidth, which is mainly because larger bandwidth leads to more
subcarrier apertures gain. Furthermore, we observe that the performance
gain of TSGE over MBWDE increases with bandwidth.

\section{\label{sec:Conclusion}Conclusion}

In this paper, we studied a delay estimation problem in the multiband
OFDM system with phase distortion factors considered. We proposed
a novel two-stage global estimation scheme that fully exploits the
multiband gains to improve the delay estimation performance. In particular,
in Stage 1, we perform a common support based sparse representation
for the coarse estimation signal model and then achieve a coarse delay
estimation using the Turbo-BI algorithm. Then, with the help of the
coarse estimation results, we further conducted a more refined delay
estimation by employing the proposed PSO-LS algorithm based on the
refined signal model in Stage 2. Finally, simulation results showed
that the proposed TSGE scheme achieves superior performance over baseline
algorithms in the InF scenario. Future work can consider a multiband
delay estimation problem for localization in a multiple-input multiple-output
(MIMO) system.

\appendix

\subsection{\label{subsec:A.-Message-Passing}Message Passing for Module B of
Turbo-BI}

The message from variable node $x_{m,l}$ to factor node $f_{m,l}$
is given by
\begin{equation}
v_{x_{m,l}\rightarrow f_{m,l}}\left(x_{m,l}\right)=\mathcal{CN}\left(x_{m,l};x_{B,m,l}^{pri},v_{B,m}^{pri}\right).\label{eq:MP1}
\end{equation}
The message from factor node $f_{m,l}$ to variable node $s_{l}$
is given by
\begin{equation}
v_{f_{m,l}\rightarrow s_{l}}\left(s_{l}\right)\wasypropto\int f_{m,l}\left(x_{m,l},s_{l}\right)v_{x_{m,l}\rightarrow f_{m,l}}\left(x_{m,l}\right)dx_{m,l}=\pi_{l}^{m}\delta\left(s_{l}-1\right)+\left(1-\pi_{l}^{m}\right)\delta\left(s_{l}\right),\label{eq:MP2}
\end{equation}
where $\pi_{l}^{m}=(1+\frac{\mathcal{C}\mathcal{N}(0;x_{B,m,l}^{pri},v_{B,m}^{pri})}{\mathcal{C}\mathcal{N}(0;x_{B,m,l}^{pri},v_{B,m}^{pri}+\sigma_{m,l}^{2})})^{-1}$.
Then the message passed from variable node $s_{l}$ to factor node
$f_{m,l}$ is given by
\begin{equation}
v_{s_{l}\rightarrow f_{m,l}}\left(s_{l}\right)\wasypropto v_{d_{l}\rightarrow s_{l}}\left(s_{l}\right)\underset{m^{\prime}\neq m}{\prod}v_{f_{m^{\prime},l}\rightarrow s_{l}}\left(s_{l}\right)=\widehat{\pi}_{l}^{m}s_{l}+\left(1-\widehat{\pi}_{l}^{m}\right)\left(1-s_{l}\right),\label{eq:MP3}
\end{equation}
where $\widehat{\pi}_{l}^{m}=\frac{p_{s}\prod_{m^{\prime}\neq m}\pi_{l}^{m^{\prime}}}{p_{s}\prod_{m^{\prime}\neq m}\pi_{l}^{m^{\prime}}+(1-p_{s})\prod_{m^{\prime}\neq m}(1-\pi_{l}^{m^{\prime}})}$.
Finally, the message passed from factor node $f_{m,l}$ to variable
node $x_{m,l}$ is given by
\begin{equation}
v_{f_{m,l}\rightarrow x_{m,l}}\left(x_{m,l}\right)\wasypropto\underset{s_{l}}{\sum}f_{m,l}\left(x_{m,l},s_{l}\right)v_{s_{l}\rightarrow f_{m,l}}\left(s_{l}\right)=\widehat{\pi}_{l}^{m}\mathcal{C}\mathcal{N}\left(x_{m,l};0,\sigma_{m,l}^{2}\right)+\left(1-\widehat{\pi}_{l}^{m}\right)\delta\left(x_{m,l}\right).\label{eq:MP4}
\end{equation}

\subsection{\label{subsec:B.-Gradient-derivation}Gradient Derivation in Turbo-BI-M
Step}

\begin{equation}
\zeta_{\delta_{m}}=-\frac{2}{\sigma_{ns}^{2}}\mathrm{Re}\left[\mathbf{b}_{m}^{\prime{}^{H}}\mathbf{b}_{m}-\mathbf{b}_{m}^{\prime}{}^{H}\mathbf{y}^{m}+\mathrm{tr}\left(\mathbf{A}_{m}\boldsymbol{\Sigma}_{m}\mathbf{A}_{m}^{H}\mathbf{S}_{m}^{H}\mathbf{S}_{m}^{\prime}\right)\right]-\frac{\delta_{m}}{\sigma_{p}^{2}},
\end{equation}

\begin{equation}
\begin{gathered}\zeta_{\Delta\tau_{l}}=-\frac{2}{\sigma_{ns}^{2}}\sum_{m=1}^{M}\mathrm{Re}\left[\left(\left(\mathbf{a}_{m}^{\prime}\left(\overline{d}_{l}+\Delta\tau_{l}\right)\right)^{H}\mathbf{S}_{m}^{H}\mathbf{S}_{m}\mathbf{a}_{m}\left(\overline{d}_{l}+\Delta\tau_{l}\right)\right)\left(\left|\mu_{l,m}\right|^{2}+\Sigma_{ll,m}\right)-\right.\\
\left.\left(\mathbf{a}_{m}^{\prime}\left(\overline{d}_{l}+\Delta\tau_{l}\right)\right)^{H}\mathbf{S}^{H}\left(\mu_{l,m}^{*}\mathbf{y}_{m,-l}-\mathbf{S}_{m}\sum_{j\neq l}\left(\sum_{jl,m}\mathbf{a}_{m}\left(\overline{d}_{l}+\Delta\tau_{j}\right)\right)\right)\right],
\end{gathered}
\end{equation}
where $\mathbf{b}_{m}=\mathbf{S}_{m}\mathbf{A}_{m}\mu_{m}$, $\mathbf{b}_{m}^{\prime}=\frac{\partial\mathbf{b}_{m}}{\partial\delta_{m}}$,
$\mathbf{y}_{m,-l}=\mathbf{y}_{m}-\mathbf{S}_{m}\sum_{j\neq l}(\mu_{j,m}\mathbf{a}_{m}(\overline{d}_{j}+\Delta\boldsymbol{\tau}_{j}))$,
$\mathbf{a}_{m}^{\prime}(\overline{d}_{l}+\Delta\boldsymbol{\tau}_{l})=\frac{d\mathbf{a}_{m}(\overline{d}_{l}+\Delta\boldsymbol{\tau}_{l})}{d\Delta\boldsymbol{\tau}_{l}}$,
$\mathbf{S}_{m}^{\prime}=\frac{d\mathbf{S}_{m}}{d\delta_{m}}$, $\mu_{l,m}$
and $\Sigma_{jl,m}$ denote the $l$-th element and the $(j,l)$-th
element of the posterior mean $\boldsymbol{\mu}_{m}$ and covariance
$\boldsymbol{\Sigma}_{m}$ associated with $p(\mathbf{x}\mid\mathbf{y},\boldsymbol{\xi})$,
which can be approximated using the $\mathbf{x}_{A,m}^{post}$ and
$\mathbf{V}_{A,m}^{post}$ calculated in the Turbo-BI-E step.

\bibliographystyle{IEEEtran}
\phantomsection\addcontentsline{toc}{section}{\refname}
\bibliography{multiband}
\end{document}